\documentclass[a4paper,fleqn]{cas-sc}

\usepackage[authoryear,longnamesfirst]{natbib}
\usepackage{multirow}
\usepackage{subcaption}

\def\tsc#1{\csdef{#1}{\textsc{\lowercase{#1}}\xspace}}
\tsc{WGM}
\tsc{QE}

\begin{document}
\let\WriteBookmarks\relax
\def\floatpagepagefraction{1}
\def\textpagefraction{.001}

\shorttitle{Rotational equivariance and locality in data-driven subgrid-scale closures}    

\shortauthors{}  

\title [mode = title]{Rotational equivariance and locality in data-driven subgrid-scale closures}  



%

\author[1]{Ryley McConkey}[orcid=0000-0003-0674-1849]
\cormark[1]
\ead{rmcconke@mit.edu}

\author[1]{Julia Balla}[orcid=0000-0001-6831-2877]
\author[1]{Elyssa Hofgard}[orcid=0000-0002-0745-9477]
\author[1]{Tess Smidt}[orcid=0000-0001-5581-5344]
\author[1]{Abigail Bodner}[orcid=0000-0003-1333-2840]

\affiliation[1]{organization={Massachusetts Institute of Technology},
            addressline={77 Massachusetts Avenue}, 
            city={Cambridge},
            postcode={02139}, 
            state={Massachusetts},
            country={United States of America}}

\cortext[1]{Corresponding author}


\begin{abstract}
Data-driven subgrid-scale closures for large eddy simulation are of significant interest in many engineering and geoscience applications. In this context, several important questions remain about the role of rotational equivariance as an inductive bias for learned tensorial mappings. We investigate whether equivariance improves accuracy, parameter efficiency, and generalization for subgrid-scale modelling at realistic filter ratios. For turbulent channel flow, we compare data-augmented non-equivariant architectures to those with equivariance as an inductive bias. We compare both pointwise and nonlocal versions of these two model classes. All models are evaluated at matched parameter counts across spatiotemporal, anisotropy, and Reynolds number generalization. We show that non-augmented models learn a small degree of equivariance directly from turbulence data, especially when that data is more isotropic. The equivariant nonlocal architecture attains the highest correlation coefficient on every generalization test at approximately half the parameter count of its non-equivariant counterpart, while the pointwise architectures do not improve on the analytical Clark baseline. Additionally, the equivariant model is more data-efficient than a non-equivariant model. The benefit of equivariance grows with the receptive field of the model, indicating that equivariance and nonlocality are both useful for the subgrid-scale closure task at realistic dataset size, parameter counts, and filter size.
\end{abstract}


\begin{highlights}
\item Architectural rotational equivariance and locality are compared for subgrid-scale closures.
\item Non-equivariant models acquire a small amount of equivariance from turbulence data alone.
\item The equivariant, nonlocal model outperforms its non-equivariant counterpart across all generalization benchmarks.
\item Pointwise models do not improve on the analytical Clark baseline.
\item Equivariance and nonlocality are both recommended for practical data-driven subgrid-scale modelling.
\end{highlights}

\begin{keywords}
Large eddy simulation \sep subgrid-scale modelling  \sep data-driven turbulence closure 
\sep equivariant neural networks
\end{keywords}

\maketitle

\section{Introduction}

In many engineering and geoscience applications, it is necessary to resolve the energy-containing scales of a fluid in order to accurately simulate turbulent flows. Large eddy simulation (LES) is a popular method for these simulations~\citep{Sagaut2006}. LES occupies the middle ground between practically unaffordable direct numerical simulation and the scale-averaging implied by Reynolds-averaged closures. However, the accuracy of LES depends directly on the subgrid-scale (SGS) stress closure, which represents the effect of unresolved scales on the resolved ones. Classical closures, including the Smagorinsky model \citep{Smagorinsky1963}, the gradient model of \citet{Clark1979}, and dynamic/mixed variants \citep{Germano1991,Vreman1996}, have been applied in LES for decades. These closures are computationally cheap and stable, but their predictive accuracy degrades in many situations required for practical turbulent flow simulations. For example, their accuracy degrades in complex flows, anisotropic grids, and during departure from equilibrium between production and dissipation~\citep{Jimenez2012,Jimenez2018}.

Increasing availability of high-fidelity DNS data has made data-driven SGS modelling a promising research direction \citep{Duraisamy2019,BruntonNoack2020,Sanderse2024}. Several model architectures have been proposed for this purpose. Pointwise multilayer perceptrons trained on the local velocity gradient tensor have been studied extensively, with applications to homogeneous isotropic turbulence \citep{Cho2024} and turbulent channel flow \citep{ParkChoi2021,Stoffer2021,Kang2023}. Nonlocal convolutional architectures that take stencils of the resolved field as input have been used to capture spatial structure that pointwise models cannot \citep{Beck2019,Maulik2019,Guan2022}. Invariant approaches reformulate the prediction in the strain-rate eigenframe to automatically enforce rotational and reflectional invariance \citep{Prakash2022,Prakash2024}. Augmentation-based approaches train on transformed copies of the DNS data to encourage the network to learn the underlying symmetries \citep{Frezat2021}. A separate line of work trains the closure inside the LES solver itself, optimizing for a posteriori statistics rather than a priori stress matching \citep{Sirignano2020,MacArt2021,Um2020,List2022}. Efforts in geophysical fluid dynamics have learned SGS closures for ocean mesoscale eddies and atmospheric convection from high-resolution simulations \citep{BoltonZanna2019,YuvalOGorman2020,Perezhogin2024,Perezhogin2025}.

When machine learning models are incorporated into physics-based simulations, it is important to respect the symmetries of the original physical system. The continuous Navier-Stokes equations are equivariant under the full rotation group $\mathrm{SO}(3)$. Filtering and discretizing the equations onto a Cartesian grid breaks this continuous symmetry to the discrete rotational octahedral group $O$, the 24 rotations that map a cube to itself \citep{Agdestein2026}. A learned SGS closure should be equivariant under this discrete symmetry group. The various data-driven SGS modelling approaches identified above address this requirement differently, if at all. Invariant models, such as those by \citet{Prakash2022,Prakash2024}, satisfy it by construction at the cost of constraining the input representation. Other forms of architectural constraints enforce equivariance through weight sharing, at the cost of additional design complexity \citep{Agdestein2026}. Outside of architectural constraints, data augmentation can be used to encourage equivariance during training~\citep{Frezat2021,Guan2023}. However, data augmentation does not guarantee exact equivariance. While the SGS community has adopted various architectures and equivariance techniques, it remains an open question as to which approach delivers the best closure for a given parameter budget, which generalizes best to unseen conditions, and whether the architectural cost of explicit equivariance is repaid by improvements in accuracy or data efficiency.

These questions around equivariance have been a major topic of research in other scientific machine learning domains. Group-equivariant convolutional networks \citep{CohenWelling2016} and steerable CNNs \citep{Weiler2019} established architectural foundations, and the geometric deep learning framework of \citet{BronsteinBruna2021} unified equivariance, invariance, and symmetry across domains. Equivariant graph neural networks for atomistic systems \citep{Satorras2021,Batzner2022} have produced state-of-the-art results in molecular property prediction and interatomic potentials, often with one or two orders of magnitude less data than non-equivariant baselines. The success of equivariant architectures in these domains has been accompanied by an active debate over when equivariance is worth the architectural cost compared with data augmentation. Empirical work has shown that equivariant networks dominate in small-data regimes but that augmented non-equivariant networks can match them as data grows \citep{Gerken2022,WangWaltersYu2022}. Theoretical analyses have shown that augmentation and equivariance produce equivalent functions in expectation under certain conditions \citep{Nordenfors2024,GerkenKessel2024,Flinth2024}, while differing in variance, training dynamics, and out-of-distribution behaviour. However, these detailed examinations of equivariance are still being explored in the context of data-driven SGS closures, despite the fluids community adopting several equivariant architectures. Recent work has begun to address the related question of how discrete and continuous symmetries interact with LES discretization \citep{Agdestein2025,Agdestein2026}, but the question of how much equivariance benefits the SGS task at realistic filter ratios remains open. Additionally, our recent work on super-resolution \citep{McConkey2026,Balla2025} shows that to some extent, training on turbulence data imparts equivariance to learned mappings, even without data augmentation.


This investigation focuses on the question of whether equivariance is worth the architectural cost for SGS modelling. We investigate three questions surrounding equivariance in data-driven SGS modelling:
\begin{enumerate}
\item Does training on turbulence data implicitly impart equivariance to non-equivariant architectures? 
\item What is the interaction between nonlocality and equivariance in a data-driven SGS model? For example, does equivariance help nonlocal constitutive closure models more?
\item At matched parameter counts, how do equivariant and non-equivariant architectures compare in accuracy, parameter efficiency, and generalization to unseen flows?
\end{enumerate}
We answer these questions through a comparison of four data-driven SGS architectures on turbulent channel flow, evaluated against a classical analytical baseline. Our primary contributions are: a matched-parameter comparison of equivariant and non-equivariant data-driven SGS architectures; quantitative evidence of implicit rotational augmentation in non-equivariant SGS networks; and cross-comparison of generalization results for equivariant/non-equivariant, and pointwise/nonlocal data-driven SGS models.

The remainder of the paper is organized as follows. Section~\ref{sec:methodology} describes the dataset, prediction task, architectures, and training protocol. Section~\ref{sec:results} presents the equivariance-error analysis, parameter-efficiency comparison, and generalization results. Section~\ref{sec:conclusion} discusses the findings and limitations.

\section{Methodology}\label{sec:methodology}

\subsection{Subgrid-scale (SGS) closure problem}\label{sec:closure_problem}

Large eddy simulation (LES) resolves the large scales of a turbulent flow and models the effect of the unresolved scales on the resolved motion. Applying a spatial filter of width $\Delta$ to the incompressible Navier-Stokes equations yields the filtered momentum equation,
\begin{equation}\label{eq:filtered_momentum}
    \frac{\partial \bar u_i}{\partial t} + \frac{\partial}{\partial x_j}(\bar u_i \bar u_j) = -\frac{1}{\rho}\frac{\partial \bar p}{\partial x_i} + \nu \frac{\partial^2 \bar u_i}{\partial x_j \partial x_j} - \frac{\partial \tau_{ij}}{\partial x_j} \ ,
\end{equation}
where $\bar u_i$ is the filtered velocity, $\bar p$ is the filtered pressure, and $\tau_{ij}$ is the SGS stress tensor,
\begin{equation}\label{eq:sgs_stress}
    \tau_{ij} = \overline{u_i u_j} - \bar u_i \bar u_j \ .
\end{equation}
The SGS stress represents the effect of the unresolved scales on the resolved scales and must be modelled. Data-driven SGS modelling aims to learn a mapping between the filtered fields and the SGS stress tensor. The exact inputs and outputs of this mapping used in the present investigation are discussed in the following.

In a similar manner to \citet{ParkChoi2021} and \citet{Prakash2022,Prakash2024}, we restrict the model input to the filtered velocity gradient tensor, $G_{ij} = \partial \bar u_i / \partial x_j$. This choice ensures Galilean invariance~\citep{Speziale1985}. While the filtered velocity vector $\bar u_i$ is not invariant under Galilean boosts, its gradient is. A Galilean invariant closure is required since the SGS stress is invariant under uniform translations of the velocity field. We do not include the filter width $\Delta$ as a separate input. As discussed below, $\Delta$ enters through the normalization of inputs and outputs.

The model output is the deviatoric SGS stress $\tau^{\mathrm{d}}_{ij} = \tau_{ij} - \frac{1}{3}\tau_{kk}\delta_{ij}$. The isotropic part can be absorbed into a modified pressure in the filtered momentum equation, and it is therefore the deviatoric part of the SGS stress which is of interest. Five components are sufficient to specify the deviatoric stress, since it is symmetric and zero-trace. We predict $(\tau^{\mathrm{d}}_{11}, \tau^{\mathrm{d}}_{22}, \tau^{\mathrm{d}}_{12}, \tau^{\mathrm{d}}_{13}, \tau^{\mathrm{d}}_{23})$, with $\tau^{\mathrm{d}}_{33} = -\tau^{\mathrm{d}}_{11} - \tau^{\mathrm{d}}_{22}$ recovered from the trace-free condition.

Inputs and outputs are normalized using the local velocity gradient magnitude. We define
\begin{equation}\label{eq:G_def}
    G = \sqrt{2 S_{ij} S_{ij} + 2 R_{ij} R_{ij}} \ ,
\end{equation}
where $S_{ij}$ and $R_{ij}$ are the symmetric and antisymmetric parts of $G_{ij}$. The normalized input and output tensors are
\begin{equation}\label{eq:normalization}
    \hat G_{ij} = \frac{G_{ij}}{G} \ , \qquad \hat\tau^{\mathrm{d}}_{ij} = \frac{\tau^{\mathrm{d}}_{ij}}{\Delta^2 G^2} \ .
\end{equation}

These normalizations are motivated by several factors. 
First, the normalizations in (\ref{eq:normalization}) render the learned mapping dimensionless, removing the explicit Reynolds number and filter width dependence from the mapping. This non-dimensionalization is a prerequisite for Reynolds number generalization. Second, the choice of $\Delta^2 G^2$ for the output scale follows the mixing-length argument underlying the Smagorinsky model~\citep{Smagorinsky1963,Pope2000}, in which the SGS stress is set by an eddy viscosity $\nu_t \sim \Delta^2 |S_{ij}|$, so that $\tau^{\mathrm{d}}_{ij} \sim \nu_t |S| \sim \Delta^2 |S_{ij}|^2$. Normalizing by $\Delta^2 G^2$ factors out this leading-order scaling and leaves the network to learn only the dimensionless departure from it. We use $G$ rather than $|S_{ij}|$ because $G$ does not vanish in regions of pure rotation and is bounded below across the dataset. This normalization is essential to the generalization tests in this study; the strain magnitude and filter width vary by orders of magnitude between the near-wall and channel-center regions and between the two Reynolds numbers. Without factoring out $\Delta^2 G^2$, a model could appear to generalize or fail to generalize purely because the output magnitude changed. The learned mapping $f$ acts between dimensionless tensors, which isolates the question of equivariance and generalization from dimensional rescaling.

 With these normalizations, the mapping that is learned is ultimately
\begin{equation}\label{eq:mapping}
    \hat\tau^{\mathrm{d}}_{ij} = f(\hat G_{ij}) \ ,
\end{equation}
where $f$ is a neural network. At evaluation time, the prediction is rescaled by $\Delta^2 G^2$ to recover the dimensional SGS stress.

\subsection{Symmetries of the discretized filtered Navier-Stokes equations}\label{sec:symmetries}
There are at least two reasons to enforce equivariance in a data-driven closure model. The first is philosophical. The Navier-Stokes equations do not depend on the orientation of the coordinate system, and the SGS stress is a tensor that transforms under rotation of the velocity field. A model that does not respect this transformation property can produce different predictions for the same physical state expressed in two coordinate systems, which is inconsistent with the physics it is meant to emulate. Enforcing equivariance removes this inconsistency by construction. The second reason is practical. A non-equivariant model must learn the transformation property from data, effectively spending parameters and training examples to discover a symmetry that is completely known in advance~\citep{CohenWelling2016,BronsteinBruna2021}. An equivariant model has this symmetry by construction, which reduces the size of the function search space~\citep{Weiler2019}. Equivariance constrains predictions on inputs that are simply new orientations of training examples not seen during training. We therefore expect equivariance to improve parameter efficiency and generalization, and the central question of this study is to quantify when and to what extent this expectation holds for SGS modelling.

The incompressible Navier-Stokes equations are equivariant under the full rotation group $\mathrm{SO}(3)$~\citep{Pope2000}. If $u_i(\boldsymbol{x},t)$ is a solution and $\mathsf{R} \in \mathrm{SO}(3)$ is a rotation matrix, then $\mathsf{R}_{ij} u_j(\mathsf{R}^{-1}\boldsymbol{x},t)$ is also a solution. The SGS stress tensor inherits this symmetry: rotating the input velocity field by $\mathsf{R}$ rotates the SGS stress tensor by the same rotation,
\begin{equation}\label{eq:continuous_equivariance}
    \tau_{ij}(\mathsf{R} \boldsymbol{x}, t; \mathsf{R}\bar{\boldsymbol{u}}) = \mathsf{R}_{ik} \mathsf{R}_{jl} \tau_{kl}(\boldsymbol{x}, t; \bar{\boldsymbol{u}}) \ .
\end{equation}
A closure mapping $f$ that predicts the SGS stress from the filtered velocity gradient should respect this transformation property. The input filtered velocity gradient and the output deviatoric SGS stress are both second-rank tensors, and transform under the same representation $D$. We say that $f$ is equivariant with respect to a group $\Gamma$ if, for all $g \in \Gamma$ and all inputs $\hat G_{ij}$,
\begin{equation}\label{eq:equivariance_def}
    f(D(g) \hat G) = D(g) f(\hat G) \ .
\end{equation}
For a rotation $\mathsf{R}$, the representation acts on a second-rank tensor as $\hat G_{ij} \mapsto \mathsf{R}_{ik} \mathsf{R}_{jl} \hat G_{kl}$, and the output transforms identically, $\hat\tau^{\mathrm{d}}_{ij} \mapsto \mathsf{R}_{ik} \mathsf{R}_{jl} \hat\tau^{\mathrm{d}}_{kl}$.

The action of $D$ described above applies to the tensor components at a single point. When the model takes spatially extended input, as the nonlocal models in this study do, the representation acts on the field in two ways simultaneously. It rotates the tensor components at every point, and it permutes the points themselves. For an input field $\hat G_{ij}(\boldsymbol{x})$ defined on the grid, a rotation $\mathsf{R}$ in the rotational octahedral group maps the field to
\begin{equation}\label{eq:field_action}
    \left[ D(\mathsf{R}) \hat G \right]_{ij}(\boldsymbol{x}) = \mathsf{R}_{ik} \mathsf{R}_{jl} \, \hat G_{kl}(\mathsf{R}^{-1}\boldsymbol{x}) \ .
\end{equation}
The argument $\mathsf{R}^{-1}\boldsymbol{x}$ relocates the value at each grid point to the rotated grid point, and the prefactor $\mathsf{R}_{ik}\mathsf{R}_{jl}$ rotates the tensor stored there. Both actions must be applied together for the transformation to be a symmetry of the discrete equations. A pointwise model is equivariant with respect to the component rotation alone, since it has no spatial extent. A nonlocal model must be equivariant with respect to the combined component rotation and spatial permutation in (\ref{eq:field_action}), a constraint enforced by the group-convolutional architecture described in Section~\ref{sec:nonlocal_models}.

The continuous incompressible Navier-Stokes equations are equivariant under the full Euclidean group in three dimensions, $\mathrm{E}(3)$. This includes translations, the proper rotations $\mathrm{SO}(3)$, and improper rotations (reflections), so that the orthogonal group $\mathrm{O}(3)$ is a symmetry of the equations. 

The relevant question for a data-driven SGS model is not which group the continuous filtered Navier-Stokes equations (Eq.~\ref{eq:filtered_momentum}) are equivariant under, but which group the discretized filtered equations are equivariant under. This distinction is often overlooked in the literature on equivariant turbulence modelling. In LES, the grid is not made fine enough for the solution to converge to the continuous solution. The discretization is part of the equation being solved, not an error to be reduced \citep{Agdestein2025}. A box filter applied to a uniform Cartesian grid defines an exact transformation of the continuous Navier-Stokes equations into a discrete set of equations \citep{Agdestein2025}. These discrete equations do not inherit the full $\mathrm{O}(3)$ symmetry of the continuous equations. They are equivariant only under the subgroup of $\mathrm{O}(3)$ that maps the grid to itself. For a uniform Cartesian grid, the proper rotations in this subgroup form the rotational octahedral group $O$, which consists of the 24 rotations that permute the coordinate axes with sign changes while preserving handedness, generated by $90^\circ$ rotations about the three coordinate axes. Including the grid-compatible reflections enlarges this to the full octahedral group $O_h$ of order 48.

In this study we enforce equivariance under the rotational octahedral group $O$, not the full group $O_h$. Extending the analysis to $O_h$ by additionally enforcing reflection equivariance is left to future work. Our prior work on implicit equivariance in turbulence focused specifically on the rotational octahedral group as the relevant discrete symmetry on a Cartesian grid~\citep{McConkey2026}, and we adopt the same group here. The equivariance condition (\ref{eq:equivariance_def}) is therefore enforced with $\Gamma = O$, where $D$ acts on a second-rank tensor as $\hat G_{ij} \mapsto \mathsf{R}_{ik} \mathsf{R}_{jl} \hat G_{kl}$ for each $\mathsf{R} \in O$, together with the spatial permutation described above when the model takes spatially extended input. This paper focuses specifically on whether enforcing equivariance under $O$ improves the accuracy, data efficiency, and generalization of data-driven SGS models.

\subsection{Models}\label{sec:models}
In this work, we implement two classes of models: pointwise and nonlocal models. The difference between these is the receptive field used to predict the deviatoric part of the SGS stress tensor. The pointwise model operates at a single point in the turbulent flow field, whereas the nonlocal model sees nearby points in addition. We implement these two classes of models to determine whether a nonlocal model is worth the additional complexity and parameter count for the SGS prediction task, as well as to investigate the utility of equivariance for both model classes.

\subsubsection{Pointwise models}\label{sec:local_models}

The pointwise models predict the deviatoric SGS stress at a point from the filtered velocity gradient at that point. They have no spatial extent and are applied independently at every grid point.

The equivariant pointwise model is the strain-rate eigenframe network of \citet{Prakash2022}. The filtered velocity gradient is decomposed into its symmetric and antisymmetric parts, and the symmetric part is diagonalized. The network input is the set of four rotational invariants $(\lambda_3, \omega \cdot v_1, \omega \cdot v_2, \omega \cdot v_3)/G$, where $\lambda_3$ is the smallest strain-rate eigenvalue, $v_1, v_2, v_3$ are the strain-rate eigenvectors, $\omega$ is the vorticity vector, and $G$ is the velocity gradient magnitude of (\ref{eq:G_def}). A fully connected, feedforward multilayer perceptron (MLP) maps these invariants to the five independent components of the deviatoric stress expressed in the strain-rate eigenframe. The predicted stress is then rotated from the eigenframe back to the prediction frame. Because the inputs are rotational invariants and the output is constructed in a frame that rotates with the input, the mapping is equivariant under the full continuous $\mathrm{SO}(3)$ rotation group by construction, thereby achieving equivariance under the rotational octahedral group. The model is also Galilean invariant, since it depends only on the velocity gradient.

The non-equivariant pointwise model is an MLP that maps the nine filtered velocity gradient components directly to the five deviatoric stress components, both in the prediction frame~\citep{ParkChoi2021,Stoffer2021}. It has no built-in equivariance. Rotational equivariance is instead encouraged during training through octahedral data augmentation, described in Section~\ref{sec:training}. The two pointwise models are matched in width and depth across the parameter sweep (see Section~\ref{sec:param_matching}), so that the primary difference between them is whether equivariance is built into the architecture or learned from data augmentation.

\subsubsection{Nonlocal models}\label{sec:nonlocal_models}

The nonlocal models predict the deviatoric SGS stress at a point from the filtered velocity gradient in a spatial neighbourhood of that point. They have a kernel size of $3\times 3 \times 3$, and operate over the $14^3$ interior field of each box. The receptive field expands with depth.

The equivariant nonlocal model is a group-equivariant convolutional network over the rotational octahedral group, implemented with steerable convolutions \citep{Weiler2019,Cesa2022}. The input field is the filtered velocity gradient and the output field is the deviatoric SGS stress. Both are decomposed into the irreducible representations (irreps) of the rotational octahedral group $O$. This group has five irreps, which we denote using the standard Mulliken notation, where $A$, $E$, and $T$ label one-, two-, and three-dimensional irreps, respectively: two one-dimensional irreps $A_1$ (trivial) and $A_2$, the two-dimensional irrep $E$, and the three-dimensional irreps $T_1$ and $T_2$. Under $O$, the trace of the velocity gradient transforms as $A_1$, its antisymmetric part transforms as $T_1$, and its symmetric traceless part decomposes as $E \oplus T_2$; the nine-component input field therefore carries $A_1 \oplus T_1 \oplus E \oplus T_2$. Although the $A_1$ component (the trace of the velocity gradient) is theoretically zero for incompressible flow, we retain it in the input field because it is often nonzero in the data owing to discretization error, and to keep the methodology extensible to compressible flows. The deviatoric SGS stress is symmetric and traceless and so carries $E \oplus T_2$. This decomposition into irreps of $O$ is performed automatically by the steerable convolution framework \citep{Cesa2022} from the specified input and output field types. The hidden layers carry features in the regular representation of $O$, which decomposes as $A_1 \oplus A_2 \oplus 2E \oplus 3T_1 \oplus 3T_2$.

Each layer is a steerable $3 \times 3 \times 3$ convolution followed by a ReLU nonlinearity, with the network depth fixed at four layers. The ReLU is applied pointwise to the regular-representation features. This preserves equivariance because the regular representation acts by permuting these features under the octahedral group, and a pointwise nonlinearity commutes with a permutation. Equivariance under the rotational octahedral group is exact and holds for the combined action of the group on the tensor components and on the spatial arrangement of the field, as described in Section~\ref{sec:symmetries}. The model is Galilean invariant, since it depends only on the velocity gradient.

The non-equivariant nonlocal model is a standard three-dimensional convolutional network~\citep{Beck2019,Guan2022}. Each hidden layer is a $3 \times 3 \times 3$ convolution followed by batch normalization and a ReLU nonlinearity, with the depth fixed at four layers to match the equivariant model. It maps the nine filtered velocity gradient components to the five deviatoric stress components, both in the prediction frame, and has no built-in invariance. Rotational equivariance is instead encouraged during training through octahedral data augmentation (Section~\ref{sec:training}). The two nonlocal models share the same depth, kernel size, and receptive field, so that the only difference between them is whether equivariance is built into the architecture or learned from augmented data. In Section~\ref{sec:param_matching} we describe the strategy used to fairly compare these various architectures.

\subsubsection{Parameter matching across architectures}\label{sec:param_matching}

The central comparison in this study is between an equivariant model and its non-equivariant counterpart at a similar parameter count. We therefore sweep model size for all four architectures and report performance against the number of trainable parameters.

The pointwise models are swept by varying the depth of the multilayer perceptron at a fixed width of 64 neurons per layer, from one to eight hidden layers. The nonlocal models are swept by varying width at a fixed depth of four layers. For the equivariant nonlocal model, width is set by the number of regular-representation copies in the hidden layers, swept over $\{1, 2, 4, 8, 16, 32, 64\}$. For the non-equivariant nonlocal model, width is set by the number of convolution channels per layer, swept over $\{4, 8, 16, 32, 64, 128, 256\}$. The sweep ranges were chosen so that the equivariant and non-equivariant models of each class span the same parameter range, allowing a direct comparison at matched parameter count.

The equivariant and non-equivariant pointwise models differ slightly in parameter count at matched depth and width, because the eigenframe model takes four invariant inputs while the non-equivariant model takes the nine velocity gradient components, which changes the size of the input layer. This difference is confined to the first layer and is minor relative to the total parameter count.

\subsection{Dataset}\label{sec:dataset}

\subsubsection{Channel flow data and box extraction}\label{sec:channel_data}

The training data and evaluation data are taken from the turbulent channel flow direct numerical simulations in the Johns Hopkins Turbulence Database (JHTDB) \citep{Li2008,Graham2016,LeeMoser2015}. We use two channel datasets, at friction Reynolds numbers $Re_\tau = 1000$ and $Re_\tau = 5200$. The $Re_\tau = 1000$ channel is the primary dataset used for training, validation, and the spatiotemporal and anisotropy generalization tests. The $Re_\tau = 5200$ channel is used as a Reynolds number generalization test case. This matches the Reynolds number generalization test in our previous work~\citep{McConkey2026}.

\begin{figure}
    \centering
    \includegraphics[width=\linewidth]{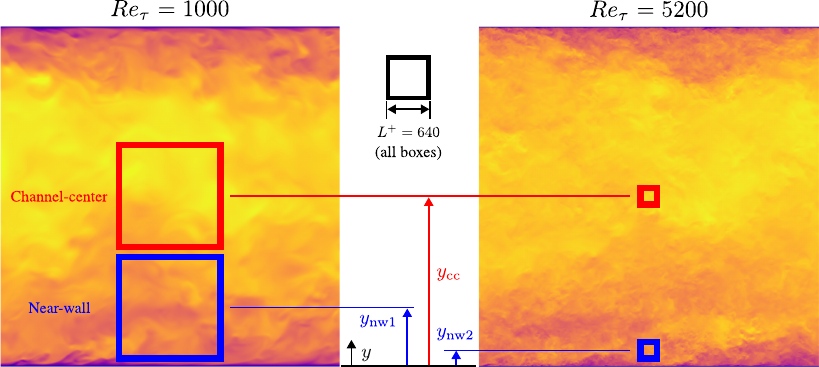}
    \caption{Extraction box locations in the two channels. Left: cross-section of the $Re_\tau = 1000$ channel. Right: cross-section of the $Re_\tau = 5200$ channel. Near-wall and channel-center boxes are drawn to scale on each channel. Center: viscous-scaled box side length, $L^+ = 640$, common to all boxes across both Reynolds numbers. The near-wall boxes are centered at a matched wall distance $y^+ = 321$ at both Reynolds numbers; the channel-center boxes are centered at the respective channel half-heights. Matching $L^+$ across Reynolds numbers keeps the filtered grid spacing $\Delta^+ = 40$ fixed.}
    \label{fig:anisotropy_y}
\end{figure}

For each Reynolds number, we extract cubic (3D) boxes from two wall-normal regions. The near-wall region is in the boundary layer, and the channel-center region is taken around the channel half-height. The near-wall boxes are centered at a wall distance of $y^+ = 321$ at both Reynolds numbers, and span $y^+ = 1$ to $641$. The channel-center boxes are centered at the respective channel half-height, giving a wall distance of $y^+ = 1000$ at $Re_\tau = 1000$ and $y^+ = 5200$ at $Re_\tau = 5200$. These two regions provide two distinct anisotropy levels at each Reynolds number. The near-wall region contains the strongly anisotropic turbulence of the boundary layer~\citep{Jimenez2012,Jimenez2018}, while the channel-center region contains turbulence closer to isotropy. The anisotropy of each region is quantified in Section~\ref{sec:anisotropy_characterization}. Figure~\ref{fig:anisotropy_y} illustrates the box locations and sizes in the two channels.

Boxes are arranged on a $15 \times 5$ grid in the streamwise and spanwise directions, over a domain of $[0, 8\pi H] \times [0, 3\pi H]$, where $H$ is the channel half-height. Each box has a fixed side length: $0.640H$ for $Re_\tau = 1000$ and $0.123H$ for $Re_\tau = 5200$. The box side length differs between Reynolds numbers so that the filter width (in viscous units) is the same for both channels, as described in Section~\ref{sec:filtering}. For $Re_\tau = 1000$, 30 evenly spaced time snapshots are extracted over the available time interval. For $Re_\tau = 5200$, 11 snapshots are extracted. Each box is queried on a uniform $64^3$ Cartesian grid using eighth-order Lagrange interpolation in space and no temporal interpolation to produce the high-resolution velocity field for that box.

Although the channel grid is anisotropic, with different spacing in the wall-normal, streamwise, and spanwise directions, each extracted box is queried on a uniform cubic grid. The symmetry analysis of Section~\ref{sec:symmetries} applies to this locally uniform cubic grid. The rotational octahedral group is the relevant rotational symmetry group for the discretized equations.

\subsubsection{Filtering and the SGS prediction task}\label{sec:filtering}

The SGS stress is generated from the high-resolution velocity field by explicit filtering. We apply a box filter with a downsampling ratio of four. Each $64^3$ high-resolution box is partitioned into non-overlapping $4 \times 4 \times 4$ blocks, and the velocity in each block is averaged to produce a $16^3$ filtered field. The filter width $\Delta$ is equal to the coarse grid spacing, which is the standard configuration for an SGS model. The high-resolution grid spacing corresponds to $\Delta x^+ \approx 10$ in viscous wall units, and the filtered grid spacing is therefore $\Delta^+ \approx 40$. These values are the same in wall units for both Reynolds numbers, which is the reason the physical box side length differs between the two channels (Section~\ref{sec:channel_data}). A filter width of $\Delta^+ \approx 40$ is representative of a typical SGS filtering task.

The SGS stress tensor is computed directly from its definition (Equation \ref{eq:sgs_stress}). The product $\overline{u_i u_j}$ is obtained by box-filtering the high-resolution velocity products, and $\bar u_i \bar u_j$ is formed from the filtered velocity. The deviatoric part is taken by subtracting the isotropic component, and five independent components $(\tau^{\mathrm{d}}_{11}, \tau^{\mathrm{d}}_{22}, \tau^{\mathrm{d}}_{12}, \tau^{\mathrm{d}}_{13}, \tau^{\mathrm{d}}_{23})$ are retained, with $\tau^{\mathrm{d}}_{33}$ recovered from the trace-free condition.

The filtered velocity gradient tensor (the model input) is computed on the $16^3$ filtered field by second-order central differences, using a grid spacing equal to the filter width. The central difference stencil is not well defined on the boundary of the filtered box, where it would reduce to a one-sided difference. We therefore discard the outermost layer of the box in every direction and retain only the $14^3$ interior block, where the velocity gradient is a true second-order central difference at every point. The SGS stress is trimmed to the same $14^3$ block, so that inputs and targets are co-located. Every velocity gradient used as a model input is therefore calculated using a second-order central difference scheme, avoiding a degraded boundary stencil. The pointwise models operate on individual points throughout this block, and the nonlocal CNN models operate on the $14^3$ box as if it were a 3D image. The filtered velocity gradient and the deviatoric SGS stress are then normalized as in (\ref{eq:normalization}), using the local velocity gradient magnitude $G$ and the filter width $\Delta$.

\subsubsection{Anisotropy characterization}\label{sec:anisotropy_characterization}

The two wall-normal regions are intended to provide two distinct anisotropy levels. We quantify the anisotropy of each region using the resolved filtered Reynolds stress. The resolved Reynolds stress is symmetric positive semidefinite; its normalized anisotropy tensor has eigenvalues confined to the barycentric triangle of \citet{Banerjee2007}.

For each region we compute the resolved Reynolds stress from the filtered velocity fluctuations. The fluctuating velocity is defined by subtracting a mean taken over the homogeneous streamwise and spanwise directions and over time within the boxes of that region. We then form the normalized anisotropy tensor
\begin{equation}\label{eq:resolved_anisotropy}
    b_{ij} = \frac{\overline{u'_i u'_j}}{\overline{u'_k u'_k}} - \frac{1}{3}\delta_{ij} \ ,
\end{equation}
and project its eigenvalues onto the barycentric triangle to obtain the barycentric coordinates $(C_{1c}, C_{2c}, C_{3c})$. These coordinates measure the proximity of the local turbulence state to the one-component, two-component, and three-component (isotropic) limiting states \citep{Banerjee2007}. A region with $C_{3c}$ near unity is close to isotropic, while a region with large $C_{1c}$ or $C_{2c}$ is strongly anisotropic.

Table~\ref{tbl:anisotropy} reports the region-averaged barycentric coordinates for each region location and Reynolds number. The near-wall regions are more anisotropic than the channel-center regions at both Reynolds numbers, confirming that the two wall-normal locations provide two distinct anisotropy levels. The channel-center regions sit close to the isotropic vertex, with $C_{3c}$ of 0.65 at $Re_\tau = 1000$ and 0.79 at $Re_\tau = 5200$. The near-wall regions are towards the one-component state, with $C_{1c}$ of 0.39 at $Re_\tau = 1000$ and 0.41 at $Re_\tau = 5200$. The two near-wall regions have similar barycentric coordinates because they are centered at a matched wall distance of $y^+ = 321$ and span the same $y^+$ range at both Reynolds numbers.

\begin{table}
\centering
\caption{Extraction geometry and region-averaged barycentric coordinates of the resolved filtered Reynolds stress anisotropy tensor for each region and Reynolds number. $\Delta^+_{\text{fine}}$ is the viscous-scaled grid spacing of the extracted $64^3$ high-resolution box and $\Delta^+_{\text{coarse}}$ is the filtered $16^3$ spacing at filter ratio~4. Larger $C_{3c}$ indicates turbulence closer to isotropy.}
\label{tbl:anisotropy}
\begin{tabular}{cccccccccc}
\hline
$Re_\tau$ & Region & $y^+_{\mathrm{min}}$ & $y^+_{\mathrm{center}}$ & $y^+_{\mathrm{max}}$ & $\Delta^+_{\text{fine}}$ & $\Delta^+_{\text{coarse}}$ & $C_{1c}$ & $C_{2c}$ & $C_{3c}$ \\ \hline
1000 & Near-wall      & 1   & 321  & 641  & 10 & 40 & 0.385 & 0.247 & 0.368 \\
1000 & Channel-center & 679 & 1000 & 1000 & 10 & 40 & 0.278 & 0.072 & 0.650 \\
5200 & Near-wall      & 1   & 321  & 641  & 10 & 40 & 0.413 & 0.281 & 0.306 \\
5200 & Channel-center & 4866 & 5200 & 5200 & 10 & 40 & 0.146 & 0.061 & 0.794 \\ \hline
\end{tabular}
\end{table}

\subsubsection{Training, validation, and test splits}\label{sec:splits}

\begin{figure}
     \centering
     \begin{subfigure}[b]{0.6\textwidth}
         \centering
         \includegraphics[width=\textwidth]{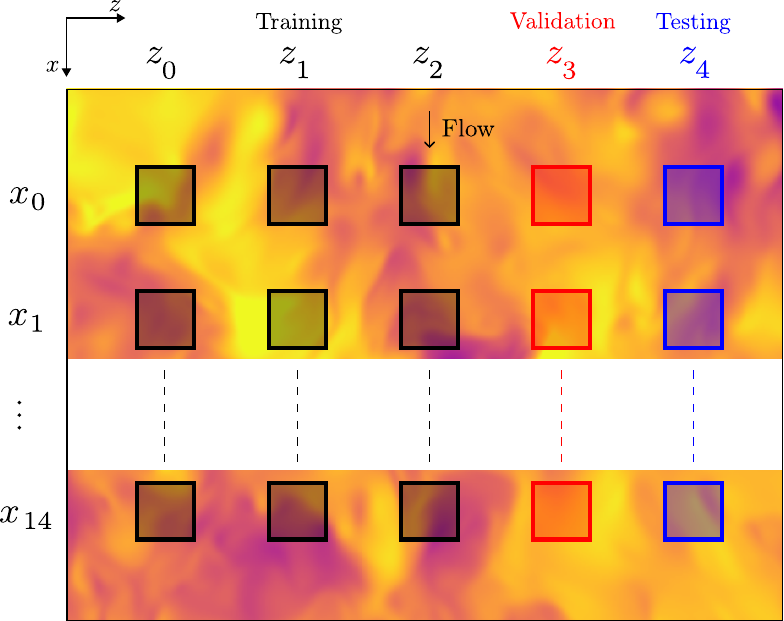}
         \caption{}
     \end{subfigure}
     \hfill
     \begin{subfigure}[b]{0.3\textwidth}
         \centering
         \includegraphics[width=\textwidth]{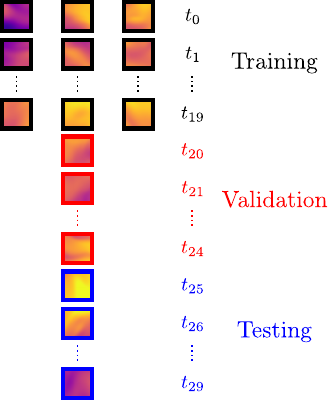}
         \caption{}
     \end{subfigure}
     \hfill
    \caption{(a) spatial and (b) temporal split of the extraction boxes into training ($z_0$, $z_1$, $z_2$), validation ($z_3$), and test ($z_4$) columns. The split is applied jointly in space and time, with no overlap between the three sets.}
    \label{fig:splits}
\end{figure}

The boxes are split in the spanwise direction with no overlap. The five spanwise columns are labelled $z_0$ through $z_4$. The model is trained on $z_0$, $z_1$, and $z_2$, validated on $z_3$, and tested on $z_4$. The 30 snapshots of the $Re_\tau = 1000$ channel are split in time over the same partition: the first 20 for training, the next 5 for validation, and the last 5 for testing. A point is in the training set only if it lies in a training column and a training snapshot, so the validation and test sets are disjoint from the training set in both space and time. Table~\ref{tbl:splits} summarizes the dataset split, and Figure~\ref{fig:splits} illustrates the locations of the streamwise and spanwise boxes.

Every trained model is evaluated on three generalization tests: spatiotemporal generalization, anisotropy level generalization, and Reynolds number generalization. Spatiotemporal generalization is measured on the held-out $z_4$ column and last 5 snapshots of the same Reynolds number and $y$-coordinate used for training. Anisotropy generalization is measured by training separate models on the near-wall and channel-center regions of the $Re_\tau = 1000$ channel, and evaluating on the other region. Reynolds number generalization is measured by evaluating models trained on the $Re_\tau = 1000$ channel on the $Re_\tau = 5200$ channel. The non-dimensionalization of the inputs and targets (Section~\ref{sec:closure_problem}) removes the explicit Reynolds number and filter width dependence, so varying the Reynolds number tests whether the dimensionless closure relationship transfers to a substantially higher Reynolds number. 
\begin{table}
\centering
\caption{Spatial and temporal partition of the extracted boxes for the $Re_\tau = 1000$ channel.}
\label{tbl:splits}
\begin{tabular}{lcccc}
\hline
Set        & Streamwise ($x$)         & Spanwise ($z$)              & Time ($t$)                  & Number of boxes \\ \hline
Training   & $x_0, \ldots, x_{14}$    & $z_0, z_1, z_2$             & $t_0, \ldots, t_{19}$       & 900   \\
Validation & $x_0, \ldots, x_{14}$    & $z_3$                       & $t_{20}, \ldots, t_{24}$    & 75    \\
Test       & $x_0, \ldots, x_{14}$    & $z_4$                       & $t_{25}, \ldots, t_{29}$    & 75    \\ \hline
\end{tabular}
\end{table}

\subsection{Training procedure}\label{sec:training}

All models are trained to minimize the mean squared error between the predicted and true normalized deviatoric SGS stress,
\begin{equation}\label{eq:loss}
    \mathcal{L} = \frac{1}{5N}\sum_{p=1}^{N} \sum_{\substack{ij \in \{11, 22, \\ 12, 13, 23\}}} \left( \tilde{\hat\tau}^{\mathrm{d}}_{ij}(p) - \hat\tau^{\mathrm{d}}_{ij}(p) \right)^2 \ ,
\end{equation}
where $\tilde{\hat\tau}^{\mathrm{d}}_{ij}$ is the model prediction, $\hat\tau^{\mathrm{d}}_{ij}$ is the true value, and the sum runs over the five independent components and all $N$ points in the training set. The loss is computed on the dimensionless tensors defined in (\ref{eq:normalization}).

The models are trained with the Adam optimizer \citep{KingmaBa2015}. The learning rate is set per architecture and held fixed across the entire size sweep for that architecture. Learning rate was chosen so that the largest model in each sweep trained stably. The batch size also varies by architecture. The pointwise models operate on the (flattened) per-point dataset, so they are trained with large batch sizes. The nonlocal models operate on full boxes and are therefore trained with smaller batches. Training proceeds for a fixed maximum number of epochs, with early stopping based on the validation loss. Here, the validation set is at the same Reynolds number and anisotropy level as the training set. The model checkpoint with the lowest validation loss is retained.

The non-equivariant models are trained with octahedral data augmentation, and the equivariant models are not. The augmentation is implemented by applying the full 24-element rotational octahedral group to every training sample, producing an augmented pool of 24 times the original size. The group action is applied jointly to the input velocity gradient field and the target SGS stress, including both the rotation of the tensor components and, for the nonlocal model, the permutation of the spatial arrangement of the field (Equation~\ref{eq:field_action}). Each epoch, a random subset of the augmented pool equal to the original training set size is drawn for training. Over the course of training this presents the non-equivariant models with the original samples under uniformly sampled octahedral rotations. This is the standard procedure by which a non-equivariant model is encouraged to learn the rotational symmetry from data, and it is the point of comparison against the equivariant models, which satisfy the symmetry by construction and require no augmentation.

A single random seed is used for each configuration. This study trains a large number of models across the four architectures and the size, dataset, and generalization sweeps, and repeating the entire set of runs over multiple seeds would multiply an already large computational cost. We therefore report single-seed results and rely on the consistency of trends across the size and dataset sweeps, rather than per-configuration variance estimates. All models were trained on a single NVIDIA H100 GPU.

\subsection{Evaluation}\label{sec:evaluation}

\subsubsection{Generalization test loss}\label{sec:test_loss}

Model accuracy is reported using the correlation coefficient $\rho_\tau$ of \citet{ParkChoi2021}, summed over the five free predicted components of the non-dimensionalized SGS stress and all evaluation points. This is the standard \textit{a priori} metric in the SGS modelling literature, and is invariant to the absolute magnitude of the target, which is important when comparing across evaluation sets with different target statistics.

\subsubsection{Equivariance error}\label{sec:equivariance_error}

The degree to which a model respects rotational octahedral symmetry is measured by the equivariance error. For an input $x$ and a group element $g$, the model is equivariant if $f(D(g) x) = D(g) f(x)$. The equivariance error measures the averaged departure from this condition,
\begin{equation}\label{eq:equiv_error}
    \mathcal{E}_{\mathrm{equiv}} = \frac{1}{|O|} \sum_{g \in O} \big\| f(D(g) x) - D(g) f(x) \big\| \ ,
\end{equation}
where $O$ is the rotational octahedral group ($|O|=24$), $D$ is the representation defined in (\ref{eq:field_action}), and the norm is taken over the output tensor components. The equivariance error is averaged over all spatial points and all samples in the evaluation set, and is reported as a single scalar for each model and each generalization set. This is the same equivariance error used in our prior work \citep{McConkey2026}. By construction, the equivariant models have zero equivariance error up to floating-point precision. For the non-equivariant models, equivariance error measures how well the rotational symmetry has been learned from octahedral data augmentation.

\subsection{Code availability}
PyTorch~\citep{Paszke2019} was used for all models and training code in this work. All of our code is available on GitHub~\citep{github_code}. 

\section{Results}\label{sec:results}
Our investigation focuses on:
\begin{enumerate}
    \item whether a non-equivariant model trained on turbulence data implicitly learns rotational equivariance~\citep{McConkey2026};
    \item whether equivariance improves parameter efficiency;
    \item whether equivariance improves generalization for pointwise and nonlocal SGS models.
\end{enumerate}
Each model is evaluated on the three generalization tests described in Section~\ref{sec:splits}, and the model size and dataset size sweeps described in Section~\ref{sec:param_matching}. 

\subsection{Equivariance error and implicit data augmentation}\label{sec:results_implicit_aug}

A non-equivariant model can only acquire equivariance by learning it from data. The equivariant models satisfy equivariance by construction and have zero equivariance error (up to floating-point precision). The question is therefore whether the non-equivariant models approach this through training, and in particular whether training on turbulence data induces approximate equivariance without explicit augmentation, a phenomenon termed \textit{implicit data augmentation}. Implicit data augmentation has been previously observed in super-resolution of turbulent velocity fields~\citep{McConkey2026}.

\begin{figure}
    \centering
    \includegraphics[width=0.8\textwidth]{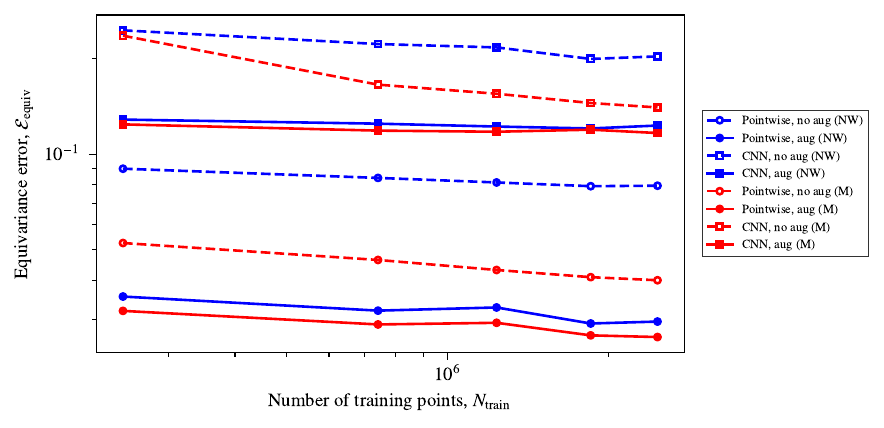}
    \caption{Equivariance error of the non-equivariant pointwise MLP and CNN, with and without octahedral data augmentation, as a function of the number of training points. Results are shown for both the near-wall (blue) and channel-center (red) regions of the $Re_\tau = 1000$ channel. Equivariance error is evaluated on the spatiotemporal validation set for each training region.}
    \label{fig:implicit_aug}
\end{figure}

Figure~\ref{fig:implicit_aug} shows the equivariance error of the non-equivariant pointwise and nonlocal models, with and without octahedral data augmentation, as a function of the number of training samples for the near-wall and channel-center regions. The non-augmented models show a decrease in equivariance error with increasing dataset size, in every region and for both model classes. This trend is the signature of implicit data augmentation. The non-equivariant models acquire some amount of equivariance directly from the turbulence data, without ever being shown explicitly rotated samples. Explicit octahedral augmentation always produces a lower equivariance error than non-augmented training at matched dataset size, but the non-augmented models have a visibly steeper slope, suggesting that the gap could close at training set sizes much larger than those considered here. The channel-center, non-augmented models reach lower equivariance error than the corresponding near-wall non-augmented models at the same dataset size. This is consistent with the more isotropic statistics of the channel-center region, which provide a wider sampling of the rotational orbit of the local turbulent state per training sample. The implicit augmentation result reported in our prior work for velocity field super resolution~\citep{McConkey2026} matches this trend. However, the overall trend here is less pronounced than the super-resolution result, which we attribute to differences in prediction task (SGS stress rather than super-resolved velocity), training split, and dataset extraction. In particular, differences in box sizes and resolution between the present work and \citet{McConkey2026} affect scales and isotropy of flow captured in each box, thereby affecting the degree of implicit data augmentation. Within each region, the pointwise non-equivariant model reaches lower equivariance error than the nonlocal CNN at matched dataset size, indicating that the pointwise mapping is easier to make approximately equivariant from finite data.

\begin{figure}
    \centering
    \includegraphics[width=\textwidth]{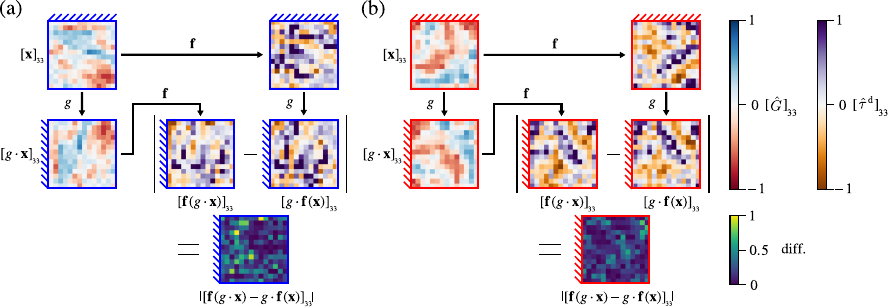}
    \caption{Commutative diagram for the non-equivariant CNNs trained without data augmentation, trained/evaluated on the near-wall (a) and channel-center (b) regions of the $Re_\tau = 1000$ channel. The input panels show the dimensionless $\hat G_{33}$ component of the velocity gradient, and the output panels show the dimensionless $\hat\tau^{\mathrm{d}}_{33}$ component of the deviatoric SGS stress, on an $xy$ slice at $z = 7$ of the $14^3$ interior. The diagram compares the path $g \cdot \mathbf{f}(\mathbf{x})$ of applying the network followed by a $90^\circ$ rotation about the $z$-axis, against the path $\mathbf{f}(g \cdot \mathbf{x})$ of applying the rotation to the input before passing through the network. The residual difference $|\mathbf{f}(g \cdot \mathbf{x}) - g \cdot \mathbf{f}(\mathbf{x})|$ quantifies the learned equivariance.}
    \label{fig:commutative}
\end{figure}

Figure~\ref{fig:commutative} demonstrates the learned equivariance of the non-equivariant nonlocal CNN trained without augmentation, by comparing the prediction of a rotated input against the rotated prediction of the original input. The two paths around the commutative diagram agree in their general spatial structure for both regions, indicating that approximate equivariance has been learned from the turbulence data alone. The residual non-equivariance shown in the difference panel is concentrated at the sharper turbulent features in the field. The difference magnitude is smaller for the channel-center region than for the near-wall region, consistent with the lower equivariance error for the channel-center non-augmented model in Figure~\ref{fig:implicit_aug}. This visualization mirrors the commutative-diagram result reported for super-resolution in our prior work~\citep{McConkey2026}, with comparable agreement and residual magnitudes.

\subsection{Parameter efficiency of equivariant architectures}\label{sec:results_param_efficiency}

We next investigate whether equivariance as an inductive bias produces a more parameter-efficient SGS model compared to learning equivariance from augmented data. We base our methodology on the scaling investigation by~\cite{Brehmer2025}. Models are compared across the model size, wall-clock time, and dataset size sweeps described in Section~\ref{sec:param_matching}.

The wall-clock time reported here is the inference time per prediction point, measured on a single NVIDIA H100 GPU. In order to achieve the fairest comparison between model architectures, we optimize the evaluation batch size separately for each model. For each model, we first identify the largest evaluation batch size that fits in memory by doubling the batch size until an out-of-memory failure is encountered. After 50 warmup iterations, we then time the forward pass for that batch size over 200 timed iterations, and divide the total elapsed time by the number of iterations and the number of prediction points per batch. Selecting the optimal batch size per architecture ensures that each model is compared at its own peak throughput on the same hardware, which is a fair point of comparison for an \textit{a priori} deployment cost.

\begin{figure}
    \centering
    \includegraphics[width=\textwidth]{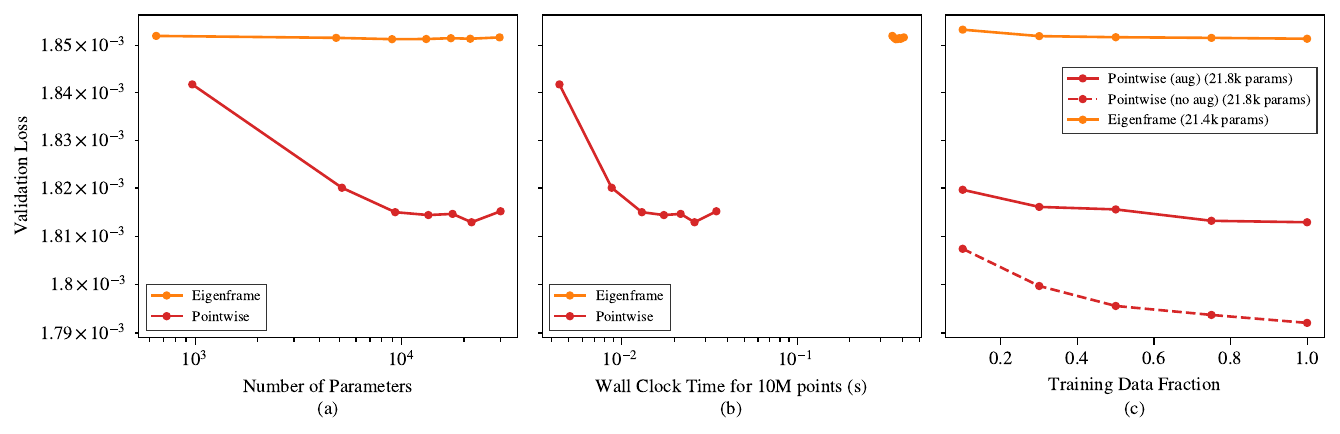}
    \caption{Validation loss for the pointwise models trained on the $Re_\tau = 1000$ near-wall dataset, as a function of (a) the number of trainable parameters, (b) the inference wall-clock time per 10M prediction points, and (c) the training data fraction. (a) and (b) show the equivariant eigenframe model and the non-equivariant pointwise multilayer perceptron trained with octahedral data augmentation. (c) additionally shows the non-equivariant pointwise model trained without augmentation. Wall-clock time is measured on a single NVIDIA H100 GPU at the optimal evaluation batch size for each architecture. The constant wall-clock time of the eigenframe model reflects the dominant cost of the strain-rate eigendecomposition, which is independent of the multilayer perceptron size.}
    \label{fig:pointwise_scaling}
\end{figure}

\begin{figure}
    \centering
    \includegraphics[width=\textwidth]{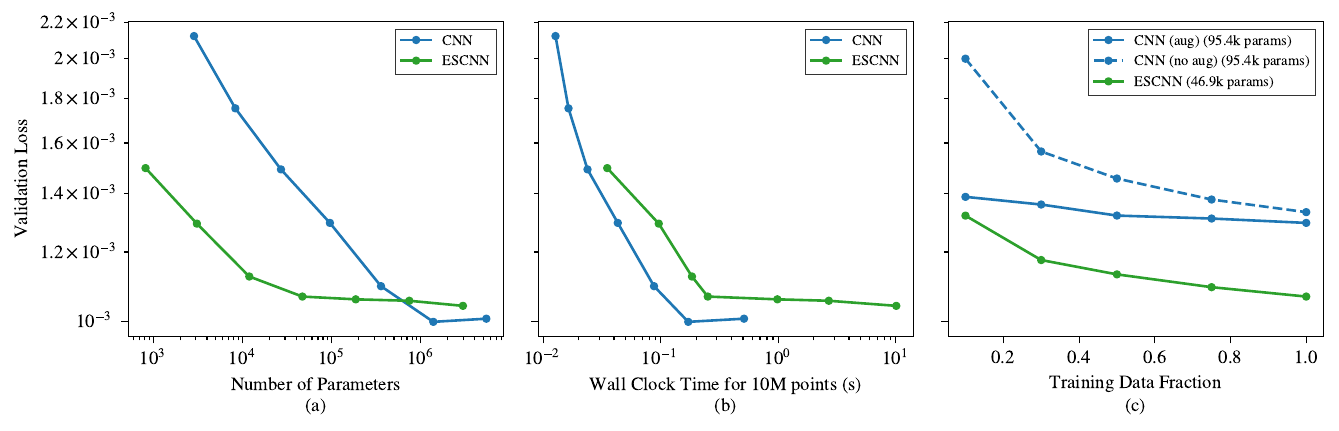}
    \caption{Validation loss for the nonlocal models trained on the $Re_\tau = 1000$ near-wall dataset, as a function of (a) the number of trainable parameters, (b) the inference wall-clock time per 10M prediction points, and (c) the training data fraction. Panels (a) and (b) show the equivariant ESCNN and the non-equivariant CNN trained with octahedral data augmentation. Panel (c) additionally shows the non-equivariant CNN trained without augmentation. The depth of both nonlocal models is fixed at four layers, as selected by the joint depth-width sweep described in Appendix~\ref{ap:hyperparameter_selection}. Wall-clock time is measured on a single NVIDIA H100 GPU at the optimal evaluation batch size for each architecture.}
    \label{fig:nonlocal_scaling}
\end{figure}

Figures~\ref{fig:pointwise_scaling} and~\ref{fig:nonlocal_scaling} show the validation loss as a function of the number of parameters, wall-clock time, and training dataset size, for the pointwise and nonlocal model pairs respectively. In the parameter scaling panel of Figure~\ref{fig:pointwise_scaling}, neither pointwise model reaches a low validation loss, and the eigenframe model's validation loss is essentially flat with parameter count. At the practical filter width selected in this work, the pointwise mapping from a single velocity gradient to the deviatoric SGS stress appears to lack the information required for accurate prediction, regardless of the parameters available to the network. In the parameter scaling panel of Figure~\ref{fig:nonlocal_scaling}, the equivariant nonlocal model (ESCNN) reaches a lower validation loss than the non-equivariant CNN at matched parameter count up to approximately $6 \times 10^5$ parameters. The CNN overtakes the ESCNN above $6 \times 10^5$ parameters, but this regime is well past the parameter counts typically used in SGS models, where the constitutive closure model must be inexpensive enough to evaluate at every grid point of a CFD solver. It is important that filter width is considered in data-driven SGS modelling investigations, as the SGS task becomes increasingly local at smaller filter widths. However, these results show that accurate SGS prediction at a realistic filter width benefits from nonlocality.

In the wall-clock time panel of Figures~\ref{fig:pointwise_scaling} and~\ref{fig:nonlocal_scaling}, equivariant models are consistently slower at matched validation loss. In the pointwise models (Figure~\ref{fig:pointwise_scaling}), the wall-clock time of the eigenframe model is dominated by the strain-rate eigendecomposition computation and is therefore approximately constant with the size of the multilayer perceptron, which is visible as a cluster of points at fixed wall-clock time in Figure~\ref{fig:pointwise_scaling}. This is a current limitation of the PyTorch implementation. While this specific part of the computation presents a bottleneck, and prevents a truly fair comparison, it should still be considered a disadvantage of the eigenframe model. For practical applications, computing millions of eigenvalues per CFD solver step in a stable and fast manner could present implementation difficulties. In the nonlocal models (Figure~\ref{fig:nonlocal_scaling}), the ESCNN is consistently slower than the CNN at matched validation loss. At matched wall-clock time, the CNN reaches a lower validation loss, but the gap is much smaller than at matched parameter count, and is smallest in the low-parameter regime. For the nonlocal models, the PyTorch convolution paths are based on highly optimized production machine learning libraries, while the steerable convolution implementation used here is less optimized research code. Nevertheless, the difficulties in optimizing equivariant computations have resulted in generally slower models at the same validation loss compared to non-equivariant models in several domains~\citep{Batzner2022}.

In the dataset size panel of Figure~\ref{fig:pointwise_scaling}, the eigenframe model maintains a higher validation loss than the augmented and non-augmented pointwise models across the entire dataset size sweep. The augmented and non-augmented non-equivariant pointwise models both improve as the dataset grows, but do not cross. The non-augmented pointwise model is in fact lower than the augmented version, in contrast to the corresponding nonlocal pair. We believe this is because the pointwise problem is too information-limited for either equivariance or augmentation to yield an advantage at the dataset sizes considered here. In the dataset size panel of Figure~\ref{fig:nonlocal_scaling}, the ESCNN attains a lower validation loss than both the augmented and non-augmented CNN across the entire range, and improves more steeply with dataset size. The non-augmented CNN approaches the augmented CNN at full training data, consistent with the implicit data augmentation result of Section~\ref{sec:results_implicit_aug}: when the dataset is large enough, training on the original samples is nearly equivalent to training on the augmented samples. The non-augmented models also have a steeper improvement with dataset size than the augmented models.

Taken together, the nonlocal results follow similar patterns reported for equivariance as an inductive bias in other domains. Equivariance is most useful in the low-parameter and low-data regime, which is exactly where SGS modelling lives in practice. A deployable closure must be small enough to run at every grid point of a CFD solver, and the training set is bounded by the cost of producing high-fidelity reference data. In this regime, the equivariant nonlocal model is strictly more parameter- and data-efficient than its non-equivariant counterpart with augmentation. Equivariance is less important at very large parameter counts and training set sizes; however, typical dataset and model sizes are well below this threshold.

\subsection{Generalization}\label{sec:results_generalization}

An important question for a data-driven SGS model is whether the learned closure generalizes beyond the flow conditions it was trained on. We evaluate generalization along the three tests defined in Section~\ref{sec:splits}: an unseen part of the spatiotemporal domain, an unseen anisotropy level, and an unseen Reynolds number. 

The Clark model is included as a non-learned analytical baseline. Clark predicts the deviatoric SGS stress from the local velocity gradient as
\begin{equation}\label{eq:clark}
    \tau^{\mathrm{Clark}}_{ij} = \frac{\Delta^2}{12} \frac{\partial \bar u_i}{\partial x_k}\frac{\partial \bar u_j}{\partial x_k},
\end{equation}
and is taken in deviatoric form to match the trained model output. The Clark model arises from a leading-order Taylor expansion of the filtered velocity field~\citep{Clark1979}, and is a pointwise mapping from the velocity gradient with no learnable parameters.

Table~\ref{tbl:generalization} reports the correlation coefficient of the five models on each of the three generalization axes, with each model evaluated in both training directions: trained on the near-wall region and trained on the channel-center region. The ESCNN attains the highest correlation on every cell of the table, regardless of training direction or generalization axis. At approximately half the parameter count, the ESCNN exceeds the augmented CNN in correlation on every cell. The two pointwise models are essentially tied across all cells, with correlation differences within 0.02 in either direction.

The anisotropy generalization gap is direction-dependent. Training on the channel-center region (less anisotropic) and testing on the near-wall region (more anisotropic) is the harder direction, and the correlation drops between the in-distribution and cross-region tests are larger than in the reverse direction. For the ESCNN, training on the channel-center yields a spatiotemporal correlation of 0.874 and a near-wall test correlation of 0.747, a drop of 0.13. Training on the near-wall yields a spatiotemporal correlation of 0.760 and a channel-center test correlation of 0.857, an improvement that reflects the lower difficulty of predicting the channel-center stress.

\begin{figure}
    \centering
    \includegraphics[width=\textwidth]{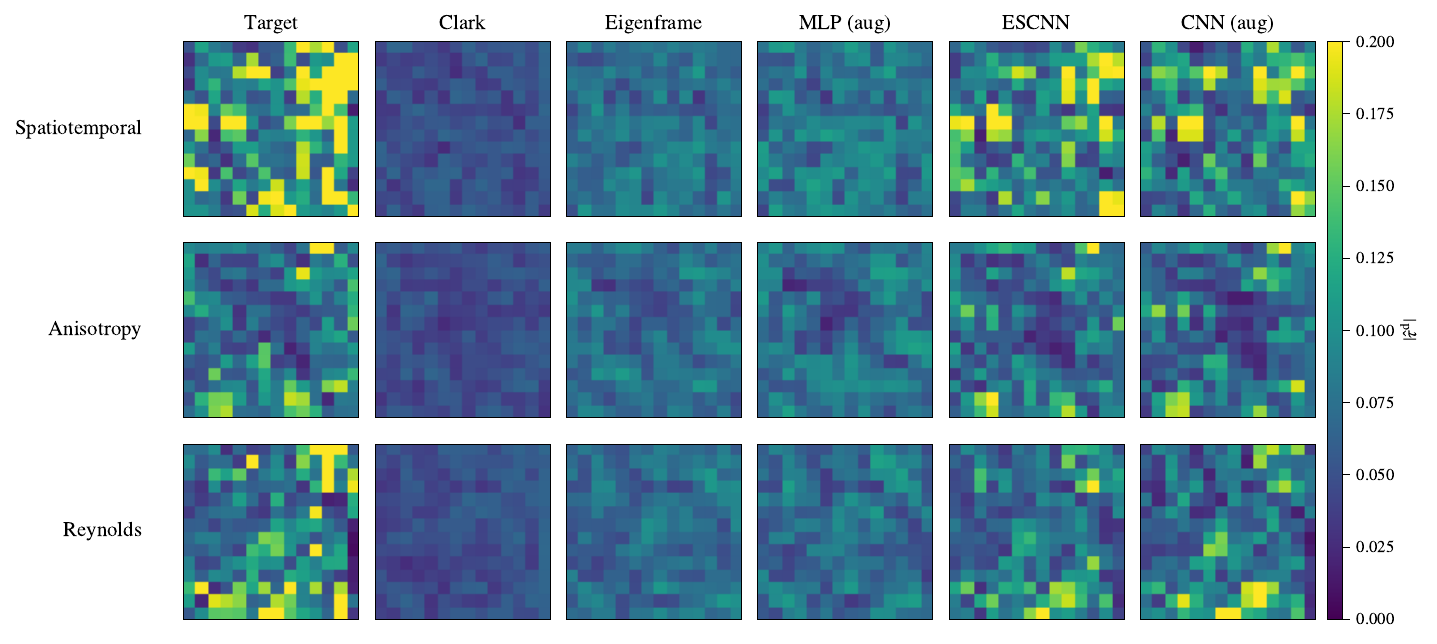}
    \caption{Magnitude of the predicted deviatoric SGS stress on an $xy$ slice at $z = 7$ of the $14^3$ interior, for the Clark model and the four trained models, compared against the filtered DNS, on the three generalization tests of Section~\ref{sec:splits}. Rows correspond to the spatiotemporal, anisotropy, and Reynolds number generalization tests. Columns correspond to the filtered DNS target, the Clark model, the Eigenframe MLP, the augmented MLP, the ESCNN, and the augmented CNN. The colourmap range is shared across columns within each row, with the maximum set to the largest magnitude attained in that row.}
    \label{fig:generalization_contours}
\end{figure}

The Reynolds number column reports correlations essentially matching the spatiotemporal column for the near-wall generalization test, and higher than the spatiotemporal column for the channel-center generalization test. For the near-wall case, the close agreement is a consequence of the dataset generation procedure. The $Re_\tau = 5200$ near-wall boxes are centered at the same viscous-scaled wall distance $y^+ = 321$ as the $Re_\tau = 1000$ near-wall boxes, and the filter width $\Delta^+ = 40$ is held constant across Reynolds numbers, so the two near-wall regions span the same $y^+$ range and have similar barycentric coordinates (Table~\ref{tbl:anisotropy}). Under these conditions the dimensionless mapping transfers cleanly across Reynolds number for every architecture considered (cf. Figure~\ref{fig:generalization_contours}). For the channel-center case, the Reynolds correlations are higher than the spatiotemporal correlations. This does not indicate a stronger form of generalization; the $Re_\tau = 5200$ channel-center region is closer to isotropy than the $Re_\tau = 1000$ channel-center region ($C_{3c} = 0.794$ versus $C_{3c} = 0.650$; Table~\ref{tbl:anisotropy}), so the deviatoric stress is easier to predict.

Figure~\ref{fig:generalization_contours} shows sample deviatoric SGS stress fields for the four trained models and the Clark model, compared against the filtered DNS data. The ESCNN tracks the spatial structure of the target most closely across all three tests, with the locations of the high- and low-magnitude regions matching those of the filtered DNS. The augmented CNN captures a similar spatial pattern, slightly less sharply than the ESCNN. The visual gap between the two nonlocal models is small, particularly on the anisotropy and Reynolds number generalization tests. The two pointwise models produce visibly similar fields to each other, and both miss the spatial structure of the target. The Clark model captures the spatial pattern reasonably well, consistent with its well-known \textit{a priori} correlation, but its predicted magnitude is substantially smaller than that of the filtered DNS. This is the standard failure mode of the Clark model and the reason it is typically used as part of a mixed model in practice~\citep{Vreman1996}. The visualization in Figure~\ref{fig:generalization_contours} is consistent with the correlation results in Table~\ref{tbl:generalization}: the model ordering is preserved across the three tests, the nonlocal models are visibly closer to the target than the pointwise models, and the equivariant nonlocal model is the best of the four trained models on each test.

The Clark model is a fixed analytical function of the local velocity gradient, so its correlation values are independent of the training direction; the table entries differ only because the spatiotemporal, anisotropy, and Reynolds number axes correspond to different evaluation regions in each direction. The Clark model attains correlations between 0.480 on the near-wall Reynolds number test and 0.818 on the $Re_\tau = 5200$ channel-center test. It trails the two pointwise trained models by no more than 0.05 in correlation on every axis and in both training directions. The pointwise models therefore do not improve substantially on the Clark baseline in terms of correlation coefficient. The nonlocal trained models substantially exceed Clark on every axis. The results in Table~\ref{tbl:generalization} combined with Figure~\ref{fig:generalization_contours} show that predicting an accurate SGS stress requires a nonlocal closure.

\begin{table}
\centering
\caption{Correlation coefficient $\rho_\tau$ \citep{ParkChoi2021} of the five models on the spatiotemporal, anisotropy, and Reynolds number generalization tests. Each model is reported in both training directions: trained on the near-wall region and trained on the channel-center region. The anisotropy column reports the cross-region evaluation; the spatiotemporal and Reynolds number columns report the evaluation at the same anisotropy level as training. Bold values indicate the best correlation in each column for each training direction.}
\label{tbl:generalization}
\begin{tabular}{llcccc}
\hline
Model & Params & Dataset & Spatiotemporal & Anisotropy & Reynolds \\ \hline
\multirow{2}{*}{Eigenframe MLP (equivariant)} & \multirow{2}{*}{21{,}445} & Near-wall      & 0.517 & 0.638 & 0.506 \\
                                              &                            & Channel-center & 0.647 & 0.509 & 0.820 \\ \hline
\multirow{2}{*}{MLP (augmented)}              & \multirow{2}{*}{21{,}765} & Near-wall      & 0.533 & 0.646 & 0.521 \\
                                              &                            & Channel-center & 0.655 & 0.521 & 0.826 \\ \hline
\multirow{2}{*}{ESCNN (equivariant)}          & \multirow{2}{*}{46{,}872} & Near-wall      & \textbf{0.760} & \textbf{0.857} & \textbf{0.751} \\
                                              &                            & Channel-center & \textbf{0.874} & \textbf{0.747} & \textbf{0.946} \\ \hline
\multirow{2}{*}{CNN (augmented)}              & \multirow{2}{*}{95{,}429} & Near-wall      & 0.700 & 0.798 & 0.690 \\
                                              &                            & Channel-center & 0.814 & 0.679 & 0.910 \\ \hline
\multirow{2}{*}{Clark model (not trainable)}                  & \multirow{2}{*}{---}      & Near-wall      & 0.491 & 0.633 & 0.480 \\
                                              &                            & Channel-center & 0.633 & 0.491 & 0.818 \\ \hline
\end{tabular}
\end{table}

\subsubsection{The importance of nonlocality}\label{sec:results_local_nonlocal}

The pointwise models predict the SGS stress from the velocity gradient at a single point, and the nonlocal models use a spatial neighbourhood. We compare the two model classes directly to determine whether additional nonlocal context improves the learned closure for channel flow at the filter ratio considered here.

The nonlocal models achieve substantially higher correlations than the pointwise models on every cell of Table~\ref{tbl:generalization}. The ESCNN exceeds the Eigenframe MLP in correlation across the three generalization axes and both training directions, despite using only roughly twice the parameters. The nonlocal advantage is present in both the near-wall and channel-center training directions, and is not larger in the more anisotropic near-wall region. The advantage of including nonlocality in the closure therefore appears to be a property of the closure problem at this filter width, rather than a property of the local turbulence anisotropy level or model architecture.

The benefit of equivariance is larger for the nonlocal class than for the pointwise class. The two pointwise models are essentially indistinguishable in correlation, while the ESCNN consistently exceeds the augmented CNN. The validation loss scaling in the parameter efficiency results of Section~\ref{sec:results_param_efficiency} shows the same asymmetry, with the eigenframe model only matching the augmented pointwise multilayer perceptron, while the ESCNN attains a substantially lower validation loss than the augmented CNN at matched parameter count. Together, these results indicate that nonlocality is necessary for the closure to capture the deviatoric SGS stress at this filter width, and therefore, the benefit of equivariance grows with the receptive field of the model.

\section{Conclusion}\label{sec:conclusion}
The discretized, filtered Navier-Stokes equations are equivariant under the rotational octahedral group, and respecting this symmetry by construction has the potential to improve the accuracy, parameter efficiency, and generalization of a learned closure. However, the advantages and disadvantages of equivariance for the SGS closure task have not been previously investigated in detail. In this study, we investigated whether enforcing rotational equivariance as an inductive bias improves a data-driven SGS closure in this context. Four architectures were compared on filtered direct numerical simulation data from turbulent channel flow at two friction Reynolds numbers. We investigated pointwise and nonlocal models. For both, we compared an equivariant model against a non-equivariant model trained with octahedral data augmentation. We evaluated each model on three generalization tests of practical relevance: spatiotemporal, anisotropy level, and Reynolds number.

Our results show that training a non-equivariant model on turbulent channel data produces a degree of equivariance without explicit augmentation, with the equivariance error decreasing as the training set grows. This implicit data augmentation effect is consistent with our prior result for super-resolution of turbulent velocity fields~\citep{McConkey2026}. We observe implicit data augmentation for both pointwise and nonlocal models in both the near-wall and channel-center regions. Explicit octahedral augmentation lowers the equivariance error at every dataset size, but the non-augmented models have a steeper slope, suggesting that the gap would close at training set sizes larger than considered here. The equivariant, nonlocal ESCNN attains a lower validation loss than the augmented CNN at every parameter count below approximately $6 \times 10^5$ parameters, which spans the practical deployment regime for SGS closures. The ESCNN attains the highest correlation coefficient on every cell of the generalization table, exceeding the augmented CNN in correlation at approximately half the parameter count. The pointwise models do not improve substantially on the analytical Clark baseline, and the nonlocal models substantially exceed both the pointwise models and Clark. The benefit of equivariance is larger for the nonlocal class than for the pointwise class, indicating that the value of equivariance grows with the receptive field of the model. Nonlocality substantially improves model predictions at the practical filter ratios considered here, which makes the receptive-field-dependent benefit of equivariance particularly relevant for SGS closures.

These findings agree with a broader pattern in scientific machine learning. The governing equations of many physical systems are equivariant under known symmetry groups, and encoding these symmetries into the architecture by construction has been shown to improve parameter efficiency, data efficiency, and generalization across a range of tasks. SGS modelling is one such task, and the present results show that the discrete rotational symmetry of the filtered Navier-Stokes equations on a uniform Cartesian grid is a useful inductive bias. We enforce equivariance under the discrete rotational octahedral group, which is the symmetry of the discretized equations on the grid, rather than the continuous $\mathrm{SO}(3)$ group of the underlying continuous equations. Matching the inductive bias to the discrete symmetry of the problem at hand, rather than to an idealized continuous symmetry that the discretization breaks, is important for SGS modelling. The closure operates on discretely filtered fields whose governing equations are equivariant under the discrete symmetry group. Augmenting with continuous rotations or enforcing $\mathrm{SO}(3)$-equivariance would impose a symmetry that the discrete equations do not satisfy~\citep{Agdestein2026}.

The practical implication is that an equivariant, nonlocal architecture is advantageous from a data- and parameter-efficiency perspective for SGS closure modelling at realistic filter ratios. The parameter- and data-efficiency gains of equivariance are concentrated in the regime where both evaluation and dataset budgets are bounded, which is the regime in which a closure is actually deployed. A closure must be inexpensive enough to evaluate at every grid point of a CFD solver at every time step, and the training set is bounded by the cost of producing high-fidelity reference data. The non-equivariant nonlocal model with augmentation matches the equivariant model only at parameter counts and dataset sizes well outside this regime. However, from an implementation and training difficulty perspective, equivariance remains a disadvantage. This recommendation matches results in other scientific machine learning domains~\citep{Brehmer2025}.

This study has several limitations. The evaluation is \textit{a priori} on filtered DNS data. The trained closures have not been tested inside an LES solver, where numerical stability, energy dissipation, and interaction with the discretization scheme determine whether a closure is usable in practice. Additionally, our investigation considers a single canonical flow type at two Reynolds numbers and one filter ratio. Generalization to free shear flows, wall-bounded flows with anisotropic grids, and different filter ratios remains to be tested. We enforce equivariance under the rotational octahedral group, and do not include reflections, which would extend the symmetry to the full octahedral group $O_h$. Future work includes evaluating the equivariant nonlocal architecture \textit{a posteriori} in an LES solver, extending the symmetry to the full octahedral group, and applying the same comparison methodology to other turbulent flows and filter ratios.
\section*{Acknowledgements}
\noindent We acknowledge the support of the National Science Foundation under Cooperative Agreement PHY-2019786 (The NSF AI Institute for Artificial Intelligence and Fundamental Interactions). J. Balla was supported by the Department of Defense (DoD) through the National Defense Science \& Engineering Graduate (NDSEG) Fellowship Program. EH was supported by the U.S. Department of
Energy, Office of Science, Office of Advanced Scientific
Computing Research, Department of Energy Computational
Science Graduate Fellowship under Award Number DESC0024386. RM was supported by the Natural Sciences and Engineering Research Council of Canada (NSERC), the Thornton Family Fund, and the MIT Electrical Engineering and Computer Science Transformation Grant. This work was supported by the U.S. Department of Energy, National Nuclear Security Administration under Award Number DE-NA0004266. This research used resources of the MIT Office of Research Computing and Data.

\section*{Declaration of generative AI and AI-assisted technologies in the manuscript preparation process}
\noindent During the preparation of this work the author(s) used generative AI in order to assist with proofreading and wording of the manuscript. The authors reviewed and edited the content as needed and take full responsibility for the content of the published article.

\appendix
\section{Hyperparameter selection}\label{ap:hyperparameter_selection}

\begin{figure}
    \centering
    \includegraphics[width=\textwidth]{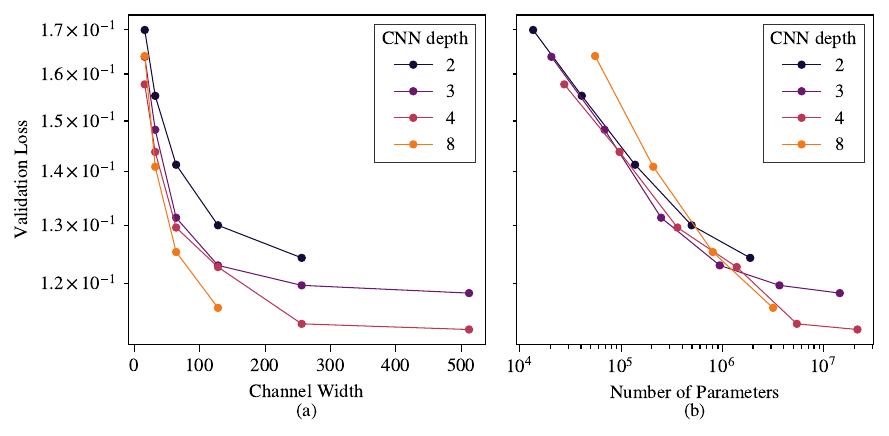}
    \caption{Validation loss on the $Re_\tau = 1000$ near-wall validation set for the non-equivariant, data-augmented CNN as a function of (a) channel width and (b) total number of parameters, with one curve per depth $\in \{2, 3, 4, 8\}$.}
    \label{fig:cnn_depth_width}
\end{figure}

\begin{figure}
    \centering
    \includegraphics[width=\textwidth]{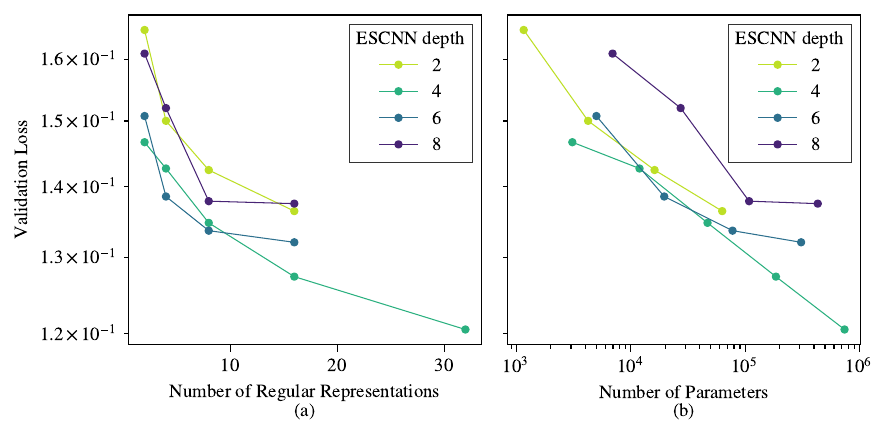}
    \caption{Validation loss on the $Re_\tau = 1000$ near-wall validation set for the equivariant ESCNN as a function of (a) number of regular representation copies per layer and (b) total number of parameters, with one curve per depth $\in \{2, 3, 4, 8\}$.}
    \label{fig:escnn_depth_width}
\end{figure}

The network depth and width of the nonlocal models were selected by a joint sweep on validation loss. For each of the equivariant and non-equivariant nonlocal architectures, we swept depth over $\{2, 3, 4, 8\}$ layers and width across the range described in Section~\ref{sec:param_matching}. All sweeps used the $Re_\tau = 1000$ near-wall training and validation sets defined in Section~\ref{sec:splits}. Figures~\ref{fig:cnn_depth_width} and \ref{fig:escnn_depth_width} show the resulting validation loss as a function of width and total parameter count, with one curve per depth. At matched channel width, deeper networks attain lower validation loss for both architectures. At matched parameter count, however, depth 4 attains the lowest validation loss over the range of parameter counts reported in the main results. Depth 4 was therefore selected as the operating point for both architectures, and the width sweep in Section~\ref{sec:param_matching} varies parameter count at fixed depth 4.

\printcredits

\bibliographystyle{cas-model2-names}

\bibliography{cas-refs}



\end{document}